# On Designing of a Low Leakage Patient-Centric Provider Network

**Yuchen Zheng[1,4], Kun Lin, Ph.D.[1], Thomas White, M.D.[3], Jeremy Pickereign[3], Gigi Yuen-Reed, Ph.D.[2]**

**AFFILIATIONS:** [1]IBM T.J. Watson Research Center, Yorktown Heights, Yorktown Height, NY, USA; [2]IBM Watson Health, Yorktown Heights, NY, USA; [3]Capital District Health Plan, Albany, NY, USA; [4]Georgia Institute of Technology, H. Milton Stewart School of Industrial and Systems Engineering, Atlanta, GA, USA

**ADDRESS CORRESPONDING TO**: Yuchen Zheng, Georgia Institute of Technology, H. Milton Stewart School of Industrial and Systems Engineering, 755 Ferst Dr. NW, Atlanta, GA 30332, [richardzyc@gatech.edu], 404-754-9191

# ABSTRACT

**Background:** When a patient in a provider network seeks services outside of their community, the community experiences a leakage. Leakage is undesirable as it typically leads to higher out-of-network cost for patient and increases barrier for care coordination, which is particularly problematic for Accountable Care Organization (ACO) as the in-network providers are financially responsible for patient quality and outcome. We aim to design a data-driven method to identify naturally occurring provider networks driven by diabetic patient choices, and understand the relationship among provider composition, patient composition, and service leakage pattern. By doing so, we learn the features of low service leakage provider networks that can be generalized to different patient population.

**Methods:** Data used for this study include personal, but fully de-identified healthcare claims acquired from Capital District Physician Health Plan (CDPHP) for diabetic patients who resided in four New York state counties (Albany, Rensselaer, Saratoga, and Schenectady) in 2014. We construct a healthcare provider network based on patients' historical medical insurance claims. A community detection algorithm is used to identify naturally occurring communities of collaborating providers. For each detected community, a profile is built using several new key measures to elucidate stakeholders of our findings. Finally, import-export analysis is conducted to benchmark their leakage pattern and identify further leakage reduction opportunity.

**Results:** The design yields six major provider communities with diverse profiles. Some communities are geographically concentrated, while others tend to draw patients with certain diabetic co-morbidities. Providers from the same healthcare institution are likely to be assigned to the same community. While most communities have high within-community utilization and spending, at 85% and 86% respectively, leakage still persists. Hence, we utilize a metric from import-export analysis to detect leakage, gaining insight on how to minimizing leakage.

**Conclusions:** We identify patient-driven provider organization by surfacing providers who share a large number of patients. By analyzing the import-export behavior of each identified community using a novel approach and profiling community patient and provider composition we understand the key features of having a balanced number of PCP and specialists and provider heterogeneity.

**Background**

Service leakage is undesirable for most healthcare provider networks. Traditionally for FFS network, out of network services correspond to less ability for payers to manage care delivery and control cost through contract negotiation and bulk purchasing. Network managers employ various strategies to minimize leakage. One common strategy is through network design, attempting to setup the most reasonable service coverage at the best fee structure as possible. Consumers are also discouraged from using out of network providers with higher copayment and coinsurance rate. As a result, FFS service network is typically large and diverse. Despite the attempt, leakage rate for specialist is still significantly at around 66.7% [12]. Similarly, the leakage rate for physicians is around 8%. [13]

In recent years, there has been an increasing emphasis on enabling value-driven healthcare, aimed at improving outcomes, lowering costs, and increasing overall access to care for patients. A prominent example of patient-centric value-driven delivery systems is the formation of provider-driven Accountable Care Organizations (ACOs). ACOs are groups of physicians, facilities and other healthcare providers, who come together voluntarily to provide coordinated high-quality care to their patients. Another example is payer-driven value network, known as narrow network, which intends to have smaller number of in-network providers but tighter control on cost and quality. The narrow network concept is gaining popularity for network targeted at specific patient populations, for example network that services specific employee population. Service leakage in these less traditional network types is even less desirable, as it undermines the premise of ability to facilitate care coordination, streamline care delivery, and manage patient outcome and quality. The implication is amplified with ACO since provider revenues are typically tied to patient outcome.

The main motivation behind our study is to characterize features of low leakage provider network. By analyzing patient past care seeking behavior we aim to surface sets of organic provider network structure, within which providers knowingly or unknowingly service the same set of patients. In prior work, Fisher et al. proposed an empirical method to organize ACOs around hospitals based on their extended medical staff [4]. Landon et al. introduced an interesting approach to identify groups of providers who might readily function as ACOs using claims data from the Medicare program [5]. We extend the latter study by considering the frequency of shared patients instead of the mere presence of patient sharing in detecting provider community to strengthen our ability in surfacing patient-driven relationships. In addition, we propose an approach to characterize network makeup and health service patterns within and across provider communities, aim to surface the design features of low-leakage network.

**METHOD**

**Data Source and Study Population:**

The medical claims data from CDPHP consist of information from each visit, the servicing provider profile, and the patient's general profile. In this study, we target the diabetic patients who were enrolled in any the commercial insurance products from CDPHP, and resided in CDPHP's four core counties in the state of New York (Albany, Rensselaer, Saratoga, and Schenectady). CDPHP has a significant market share in the region and contracts with majority of healthcare service providers of varying sizes in the area. All products are included to maximize the set of providers analyzed. Diabetes is one of the most prevalent chronic conditions in the U.S. and can often be managed with proper ambulatory care services and treatment adherence. We studied Type 1 NPIs, who are listed as individual healthcare providers, to focus on understanding patient-driven provider seeking patterns. Each provider rendered service to at least five unique diabetic patients in order to be included in the analysis. The connection between two providers is dropped if they only have one shared patient, so that only the major contributors to the provider network are included.

**Translating claims data to provider network**

Based on the chronological order of each patient's utilization sequence, we capture the connection between each patient and providers visited. Since each provider visit can result in multiple lines of claim, we define a patient visit to a provider based on the start date of the claim. We extract patient ID, NPI, and the date of the visit from each row of the claims data. A patient visited a pair of providers during the study period is considered as a "shared patient". The links are weighted by the number of patients shared by the providers. That is, the more patients two providers shared, the tighter the connection between them. From there, we are able to translate patient-level claims into a network of providers that describes the realization of patient sharing. Each node of the network represents a provider, and each weighted edge connecting nodes indicates the frequency of patient sharing.

**Patient-centric provider network**

The large cohort provider network diagram displays the overall relationship among providers but fails to unveil the interesting details of how each individual or groups of providers behave. The proposed patient-centric network design utilizes a modularity maximization algorithm develop by Newman [6] to detect provider communities, where providers in the same community are more connected than a random graph with similar characteristics. The algorithm partitions the entire network into sub-networks such that providers are more likely to share patients within the communities than across communities. The communities are non-overlapping and consist of distinct groups of providers. The network analysis is implemented in R with *igraph*. The graphs are rendered with Gephi, which utilizes the Fruchterman-Reingold algorithm to optimally position the providers (nodes) such that the connection between two providers (edge) in a crossing-free fashion [7].

**Patient assignment**

Patients are assigned to each community based on plurality. In other words, if a patient visited providers from a specific community more than providers from a different community, we assign the patient to the former community. The within community utilization for each patient is defined as the percentage of visits that happened inside the assigned community. Similarly, within community spending is defined as the percentage of spending inside the assigned community. Assigning patients by plurality naturally maximizes the within community utilization and better describes the patient population that was served by the providers in each community. Another patient assignment scheme assigns patients to their imputed Primary Care Provider (PCP) community. The imputed PCP for each patient is determined by CDPHP based on the plurality of his/her primary care visits over the previous 12 months. The ACO design suggested by Medicare requires participating ACOs to include a sufficient number of primary care ACO professionals for the number of Medicare beneficiaries assigned to the ACO [8].

**Characteristics of the detected network**

We calculate the modularity score, which measures how unlikely the detected communities are formed randomly, as an indication of how connected the detect network is. We then explore provider level information, namely provider's affiliation with healthcare organizations based on their billing tax code and their servicing specialties. This allows us to estimate the PCP-to-Specialty ratio in a given community and explore the potential impact of the overarching structure from established health systems and provider groups. We estimate Herfindahl Index for each community is calculated to measure whether a community is dominated by a few big organizations, or composed of many smaller ones. The index is the sum of squares of the organizations' market share, in terms of number of providers affiliated, and ranges from 0 to 1. A value near zero indicates a detected community with many small organizations whereas a value of 1 indicates a single monopolistic organization. Patient level information, including home county, claim cost, risk level, comorbidities and etc. are also helpful in describing the characteristics of each provider community.

**Import and Export of Services**

The detected naturally occurring provider communities differ on their patients and servicing specialties. Some tend to have more PCPs, while others tend to have more specialists. Since the services available in each provider community may not perfectly match the demands of the community, some communities are not self-sufficient. This mismatch is manifested with patients going out-of-community for services, which is undesirable in patient-centric network management. We define the service provided to a patient by providers outside his/her own community as "an import of service" to the patient's community. When a physician provides a service to patients from another community, we consider it as "an export of service". Here, the service is defined by the provider's servicing specialty, such as Family Medicine, Endocrinology, and Ophthalmology. If a community is importing a significant amount of services from other communities, substantial leakage has occurred and suggests the needs to fill the specific specialty gaps. If a community is exporting a significant amount of services to multiple communities, it suggests opportunities to establish those specialties as common services for all networks. The import-export analysis facilitates the understanding of community specialization and inter-community specialty dependencies. We conduct a trade flow analysis amongst the provider community for the most traded specialties to surface the complex web of import and export relationships. In addition, we quantify this kind of relationship by the Revealed Comparative Advantage (RCA) Index, which is broadly used in international economics for calculating the relative advantage or disadvantage of a certain country in a certain class of goods or services as evidenced by trade flows [9].

$$RCA(c,i) = \frac{\frac{x(c,i)}{\sum_i x(c,i)}}{\frac{\sum_c x(c,i)}{\sum_i \sum_c x(c,i)}}$$

Where x(c,i) is the number of service counts that community c imports/exports service specialty i.

When the RCA(c,i) with regards to export is greater than 1, it indicates that community c is efficient at exporting service specialty i. If the RCA (c,i) with regards to import is greater than 1, community c is in great need of that service. The RCA index is calculated for each community-specialty pair, both for import and export of services.

**RESULTS**

After applying the filters where a provider has to serve at least five distinct diabetic patients, and two providers have to share more than one patient in order to establish a connection, our study population has 2,213 different individual providers and 9,651 diabetic patients. Detected communities with fewer than 50 providers are excluded, resulting in six major communities with on average 347 providers per community. Although the exclusion eliminated 38 small provider communities of size 3.5 on average, it only removed 6% of the providers (133) and 7% of patients (696). Providers in these smaller communities have lower utilization and more likely to be out-of-state. The final modularity score for the detected network is 0.51 which indicates a strong community structure with sufficiently high level of inter-network patient sharing.

Looking at the lowest leakage communities (2 and 5), measure both by % in-community spending and utilization, we can see that there is a "sweet-spot" of PCP-to-specialist ratio at around 45% that allows these communities to be reasonably self-sufficient in serving average risk patient pool. For communities with under or over PCP-to-specialist ratios, the % in-community spending and utilization tend to suffer. Furthermore, when the community has a high Herfindahl index, the % in-community spending and utilization tend to be lower. Communities 4 and 6 have proportionally more few large provider organizations than smaller provider groups and have higher leakage rates. Further investigation show that the large provider organizations are formed around specific specialty services as opposed to general hospital systems that offer comprehensive care. For instance, the top organizations in Community 4 are cardiovascular-centric health systems, whereas the top organizations in Community 6 are mostly primary care provider organizations. On the contrary, Community 5 with the least leakage rate, the top organizations are regional multi-specialty hospital systems.

The four core counties in the study are geographically adjacent to each other and are well connected by highways. Figure 2 illustrates the geographical distribution of the patients served by the detected provider communities. Almost every community has an "anchoring" county where majority of the patients reside. For example, Community 2 has seventy percent (70%) of the patients reside in Schenectady county and Community 5 has ninety six percent (96%) of patients residing in Saratoga County. Nevertheless, other provider communities serve patients across multiple counties despite geographical distance. For example, Community 4 has 50% of the patients residing in Albany County, 21% in Rensselaer County and 14% in Saratoga; while Community 6 is a small community consisting of mostly PCPs, with 48% of patients residing in Albany County and the rest evenly distributed among other counties.

To examine the between-community trade pattern, Figure 3 illustrates the trade flow of the top 4 specialties based on import/export volume. The percentage shown on the edge is the percent of a specific specialty been imported/exported between two community pairs. Communities 3 and 6 are the primary importers of cardiovascular-related services, while Community 4 is the primary exporter. Community 1 stands out with approximately the same amount of import and export of cardiovascular services. From a volume perspective, there is an opportunity for Community 1 to reduce leakage by shifting the demand from out-of-community providers to within-community providers.

The RCA analysis surfaces communities who are relatively more effectively in importing/exporting certain specialties to the overall provider pool, as shown in Table 2. The usage of RCA score takes the total volume of import and export for each community into consideration, which avoids biasing towards the big communities. The largest overall imported and exported specialty is Cardiovascular Diseases, where most of the exports originated from community 4. Considering that community 4 has a large amount of total export, a RCA score of 1.9 is very significant. Community 5 has an import of Gastroenterology with RCA score of 65 because Gastroeneterology is among the only very few imports to community 5.

## DISCUSSION

## APPLICATION TO LOW LEAKAGE PROVIDER NETWORK FORMATION

Our analysis of naturally occurring provider community surfaces some interesting relationships between leakage, network structure, and patient risk profiles and geography. Provider communities with extreme patient risk profiles tend to have the lowest within-community utilization (i.e., highest leakage). Specifically, Community 4 is the highest leakage community, which has the lowest PCP-specialist ratio and attracts the highest risk pool. The second highest leakage community is 6, which has the highest PCP-Specialist ratio and attracts the lowest risk pool.

It is noteworthy that the PCP-specialist ratio varies amongst provider communities attracting average patient risk pools (CHCC scores between 4 and 6 for diabetic patients). Communities 2 and 5 attract patients primarily from a single county (Schenectady County and Saratoga County, respectively) and manage to achieve the highest within-network utilization of over 90% under a roughly 44% PCP-specialist ratio. On the other hand, the PCP-specialist ratio of Communities 1 and 3 is 31% and 56%, respectively, but both attract patients across Albany County and Rensselaer County and have lower within community utilization rates of around 80%. This suggests a potential "sweet spot" of around 45% PCP-Specialist ratio to support low leakage.

In addition, provider community formed around large provider organizations, who has higher Herfindahl score, is at higher risk of larger service leakage. This finding can be counter-intuitive yet is particularly relevant given the recent trend of merger and acquisition amongst provider organizations. Large organization that focuses on specific services is driving most of the leakage. Network designer needs to look beyond organizational size, but in the mix of specialty offered in the organization.

Import-export analysis enables network designer to identify areas of leakage risk and the next-best service specialty to add to a network. Looking at sheer volume, certain communities should be self-sufficient in certain specialties. However, the trade flow analysis demonstrates how certain communities import and export almost the same amount of certain specialty service, for example cardiovascular services and Ophthalmology services in Community 1. Given that chronic disease patients tend to have long-term relationship with their specialty providers, without considering this inter-community trade can lead to high patient dissatisfaction when setting up new ACO or narrow network. This insight allows network designer to properly mitigate such leakage risk with proper contractual relationship.

Our RCA analysis also shows clear surplus and deficit of servicing specialties in different provider communities. Generally, communities with a surplus of a certain specialty will export and communities with a deficit of certain specialties will import to balance their needs. In our study, Communities 1 and 4, both lack PCP related specialties such as Family Medicine and Internal Medicine, have to import from Community 6 which is made up of mostly PCPs. The medical college where many providers in Community 4 are affiliated with is well-known for its Cardiovascular Science program, resulting in a good amount of cardiovascular specialty exported from Community 4. By studying the dynamics across communities, we are able to identify the services that can be best provided by providers outside of the contract community which may call for special contracting considerations that target at meeting the expect service leakages.

In addition, the import/export analysis will also help identify the next-best servicing specialty to add to any community to further minimize service leakages. For example, Internal Medicine and Family Medicine providers should be added to community 4, and more diverse specialties should be included in community 6, to reduce the leakage rate. An ideal patient-centric network should be self-sustainable, and the steps layout above can contribute to the construction of such.

**OPPORTUNITIES TO INFORM GENERAL PATIENT-CENTRIC NETWORK DESIGN**

In addition to characterizing a low leakage community, our approach also highlights some of the opportunities and risks we face when forming patient-centric value network like narrow-network and ACOs. While we only consider Type 1 providers, we find that providers in most detected communities are affiliated with one or a set of provider institutions, such as general hospitals, medical schools, and healthcare groups. In general, providers contracted to the same institutions are more likely to work with each other. Out of the top 20 organizations with the highest number of affiliated providers, 16 of them have at least 75% of the affiliated providers belonging to the same community. There is less contractual and administrative burden to contract at the institutions level instead of at the individual provider level. By considering provider communities with respect to provider affiliations, we are able to study different potential scenarios for network formation. The two scenarios are: 1) a community is dominated by a few large provider organizations; 2) a community is a collection of many smaller providers that are not legally affiliated with each other. The Herfindahl Index is a good measurement that can be used to determine the characteristics of the institutions and set up opportunities for potential collaborations. Communities with a low Herfindahl Index, such as Communities 1, 2, 3, and 5 have many small organizations which can be good candidates for a third party like an insurance payer to facilitate narrow network formation or shared savings programs. Communities 4 and 6 with Herfindahl

Index of 0.27 and 0.18 are dominated by a few large provider institutions which may indicate opportunities for provider-led ACO.

Although we only focus on the diabetic patient population in our study, there is still much patient heterogeneity, even within the same community, thus patient composition is an important factor in ACO formation. In addition, geography also plays an important role in ACO formation. Therefore, different ACO formation strategies are needed based on both geography and patient composition. In geographically isolated communities such as Community 5, which has weaker connections with other communities due to the additional cost of seeking out-of-community care, patients are uniformly distributed across different health risk profiles (Figure 2). In this case, becoming self-sufficient should be a top consideration during the ACO formation process. On the other hand, Communities 1, 3, 4 and 6 are well connected and patients from these communities generally have more pronounced health risk profiles as evidenced by Figure 2. For example, Community 6 has the healthiest patient population and consists mostly of PCPs rendering their services both inside and outside of the community. Community 4, which specializes in severe conditions such as cardiovascular diseases, has the least healthy patient population. This information is valuable during new patient enrollment where a health plan can recommend the new patient to join a specific ACO based on his/her health risk profile (i.e., patient cluster) and geography. In summary, the patient segment analysis we propose enumerates the patient composition of each community, informing us about ACO contractual needs and care opportunities.

Overall, the proposed methods offers a systematic way to effectively surface provider networks that serve a specific pool of patient population; hence providing insights to network designer that would not be available through other standard summary methods. The community detection algorithm unveils not only first order connection among providers, where two providers are connected directly through a patient share instance, but also higher order connections that are very likely to happen. Since the detection method only utilizes patient visit patterns, it gives us the luxury of using other provider and patient level information to gain more insights on the resulting leakage patterns and network structure.

**LIMITATIONS OF THE ANALYSIS**

We recognize that our study has several limitations. We conduct the analysis using only 2014 data which does not capture network changes over time. Further analysis can be done using longitudinal data set to incorporate the temporal aspect of the system. The study population only consists of four counties in the state of New York, and the geographical information is limited to the county level. More granular location data with better resolution would further help identify the relationship between the network formation and geographical locations. In addition, some communities are too big to be contracted directly. In this case, geographically isolated communities can be subdivided to form smaller provider networks that have specific target patient populations based on their health conditions. These data could also be used to develop quality related measures, which would be helpful in determining whether there are discrepancies in service quality among communities. Incentives can be provided to improve the underperforming networks and help patients to choose the most suitable ones.

**CONCLUSION**

Using our proposed approach, we have observed the following desirable properties of minimum leakage network in our data: 1) a PCP-to-specialist ratio of about 45% and 2) provider heterogeneity (typically supported by many smaller providers or multi-specialty organizations. However, even if one were to follow these guidelines strictly, they might still face many practical constraints, in which case the import-export analysis presented in this paper will offer guidance on leakage risk mitigation and possible collaboration/expansion in network design.

# Declaration

**Ethics Approval and Consent to Participate:** The Manuscript reporting studies does not involve human participants, human data and/or human tissues.

**Consent for Publication:** Not applicable.

**Availability of Data and Material:** The datasets generated and analyzed during current study are not publicly available since the data used are proprietary information from CDPHP.

**Financial Disclosure Statement:** The authors have no financial relationships relevant to this article to disclose.

**Conflict of Interest Statement:** The authors have no conflicts of interest to disclose.

**Contributors:**

Yuchen Zheng has developed and implemented the methodology in this project. He has performed interpretation of data, produced the outputs included in this paper, and has been primarily responsible in drafting this article. Dr. Kun Lin has assisted in the data analysis and interpretation of the results, and he has contributed in the implementation of the methodology of this paper. He also participated in drafting this article. Thomas White and Jeremy Pickereign have provided input on knowledge about claims data, and they have assisted with interpretation and relevance of the results derived in this paper. Dr. Gigi Yuen-Reed conceptualized this research study, assisted in the data analysis and acquisition of data. She is ultimately responsible for overseeing the data analysis and manuscript preparation.

**Acknowledgments:**

The authors are thankful to Jianying Hu for assisting and supporting the project. The authors are thankful to David Jeans for providing helpful feedback.

| Community | 1 | 2 | 3 | 4 | 5 | 6 |
| --- | --- | --- | --- | --- | --- | --- |
| # of NPI | 622 | 452 | 344 | 326 | 224 | 112 |
| # of Patients | 3,069 | 1,948 | 2,106 | 605 | 760 | 467 |
| PCP-Specialist Ratio | 31.00% | 43.60% | 56.30% | 19.70% | 44.90% | 85.50% |
| % within utilization | 85.30% | 91.00% | 83.00% | 69.50% | 99.40% | 71.00% |
| % within spend | 83.30% | 90.00% | 83.30% | 66.70% | 97.90% | 75.00% |
| Herfindahl Index | 0.04 | 0.07 | 0.05 | 0.27 | 0.07 | 0.18 |
| PMPM | $1,536 | $1,359 | $1,216 | $3,586 | $1,364 | $1,008 |
| CHCC Risk Score | 5.4 | 5.9 | 4.7 | 10.2 | 4.6 | 3.2 |
| Risk-Adjusted PMPM | $284.44 | $230.34 | $258.72 | $351.57 | $296.52 | $315.00 |

TABLE 1: *Key patient and community structure characteristics for each provider community.*

| Import | | | Export | | |
|---|---|---|---|---|---|
| Community | Specialty | RCA | Community | Specialty | RCA |
| 1 | Dermatology | 2.3 | 1 | Endocrinology | 2.1 |
| 1 | Family Medicine | 1.2 | 1 | Ophthalmology | 1.5 |
| 2 | Endocrinology | 3.3 | 2 | Dermatology | 4.9 |
| 2 | Internal Medicine | 1.1 | 2 | Family Medicine | 1.7 |
| 3 | Ophthalmology | 1.4 | 3 | Family Medicine | 1.7 |
| 3 | Cardiovascular Disease | 1.1 | 3 | Ophthalmology | 1.5 |
| 4 | Internal Medicine | 1.8 | 4 | Vascular Surgery | 4.9 |
| 4 | Family Medicine | 1.7 | 4 | Cardiovascular Disease | 1.9 |
| 5 | Gastroenterology | 65.0 | 5 | Family Medicine | 6.3 |
| 6 | Cardiovascular Disease | 6.2 | 6 | Family Medicine | 2.6 |

TABLE 2: *Top imported and exported specialties for each provider community.*

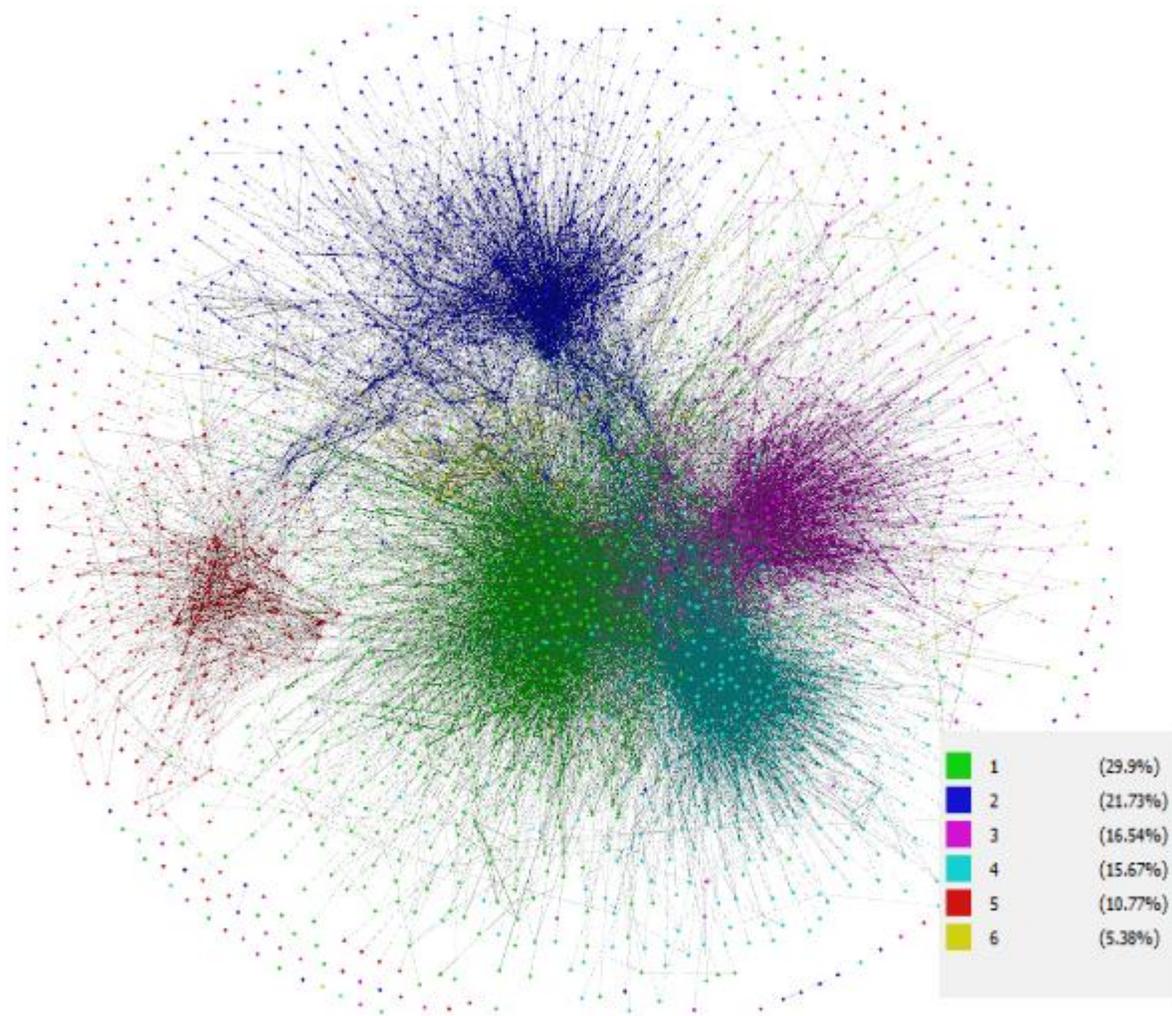

FIGURE 1: *Network structure color coded by community IDs. Nodes are providers and edges are instances of patient sharing. The edges with a weight less than 5 are discarded from the figure for ease of displaying.*

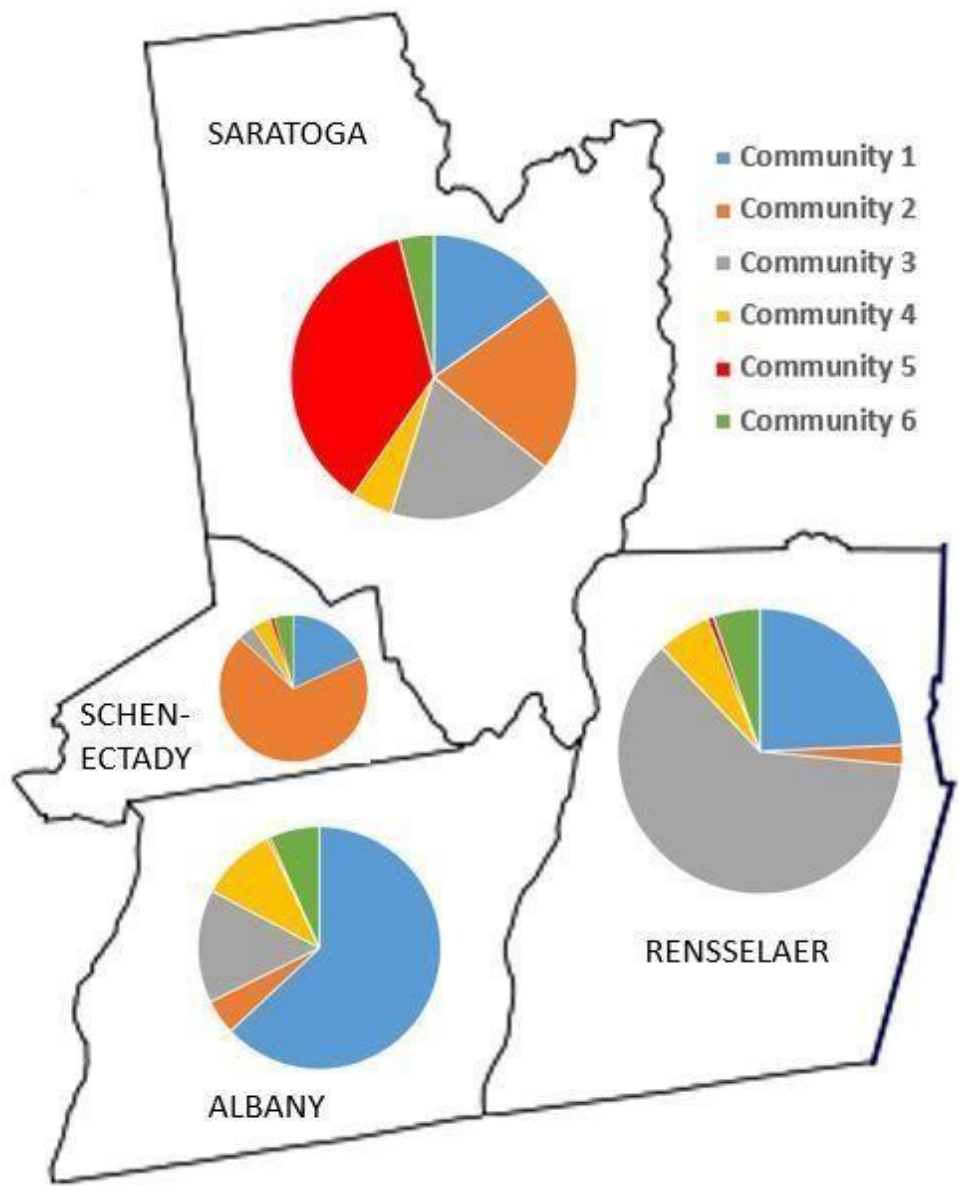

FIGURE 2: The map displays the four counties targeted in the study: Albany, Rensselaer, Schenectady and Saratoga along with the distribution of the patient population from each community.

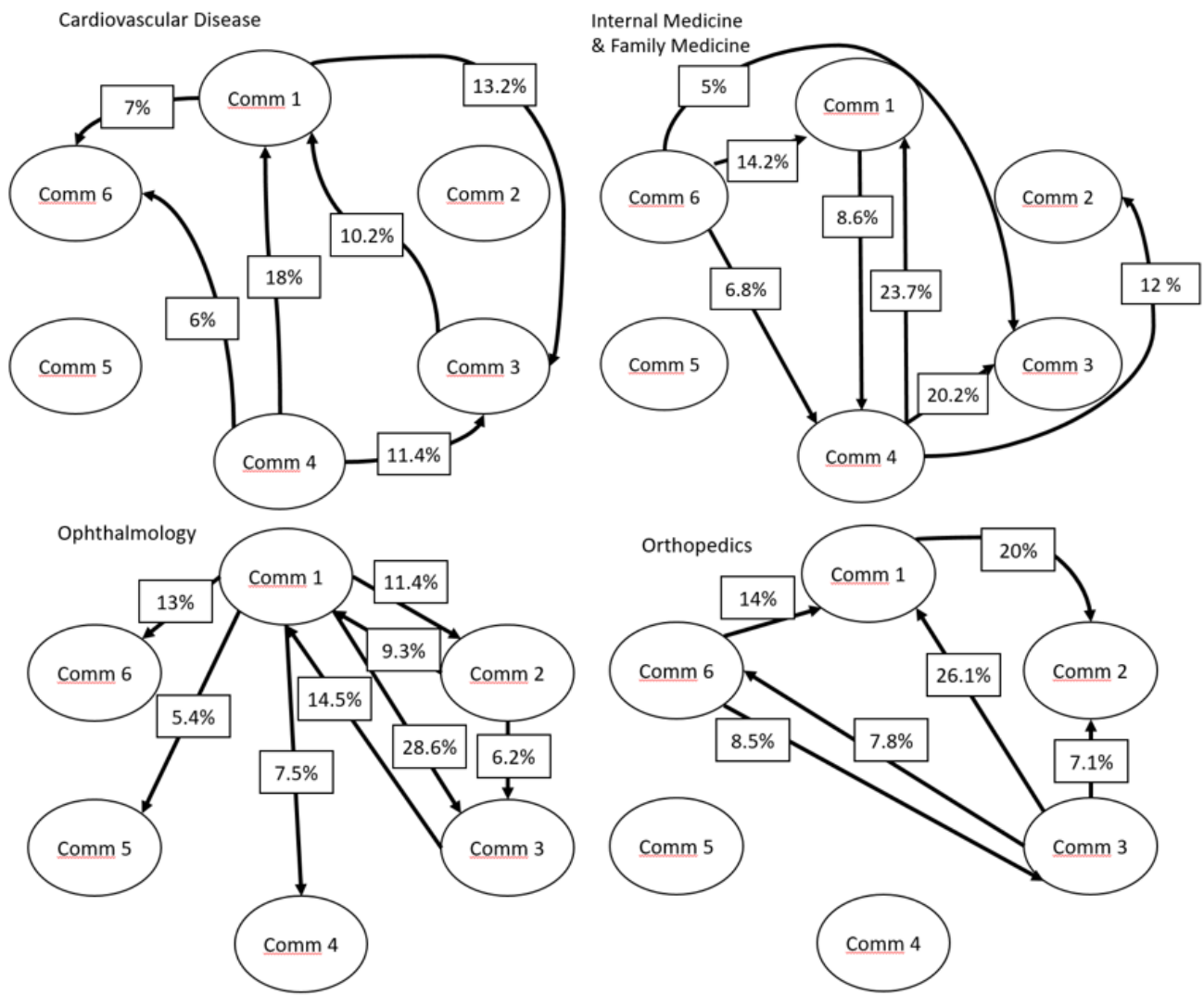

FIGURE 3: The import/export patterns of the top four imported/exported provider specialties by visit volume. The proportion of certain specialty being imported/exported for each community pairs is displayed as percentage on the edge. Edges with more than 5% are shown.